\documentclass[aip,pop,groupedaddress,preprint]{revtex4-1}
\usepackage{physics}
\usepackage{amsmath}
\usepackage{graphicx}
\usepackage{subfigure}
\usepackage{hyperref}
\usepackage{bm}
\begin{document}
\title{Different effects of suprathermal electrons and ions on drift instabilities in non-uniform plasmas} 
\author{Ran Guo}
\thanks{Author to whom correspondence should be addressed}
\email{rguo@cauc.edu.cn}
\affiliation{Department of Physics, College of Science, Civil Aviation University of China, Tianjin 300300, China}
\begin{abstract}
The paper studies the drift instability in Kappa-distributed inhomogeneous plasmas. 
Weak inhomogeneity and local approximation are assumed.
The linear dispersion relation is expressed in a novel integral representation that involves the standard plasma dispersion functions defined in Maxwellian plasmas rather than the generalized plasma dispersion function modified by the Kappa distribution.
The analytical solutions of wave frequency and growth rate are derived when the wave speed is larger than the ion thermal speed but smaller than the electron thermal speed.
The unstable drift mode is found to have a lower limit of wavenumber due to the suprathermal electrons,
which does not exist in Maxwellian plasmas.
The population of suprathermal ions affects the critical wavenumber for instability but does not determine its emergence.
Moreover, the wave frequency, growth rate, and critical wavenumber are numerically solved from the dispersion relation.
The results suggest that the critical wavenumber cannot be neglected with the typical parameters of space plasmas.

\end{abstract}
\maketitle

\section{Introduction}
\label{sec:intro}
The drift wave is one of the essential physical phenomena in inhomogeneous plasmas.
It finds various applications in both laboratory and space plasmas. \cite{Horton1999}
In astrophysical plasmas, 
Ilyasov \textit{et al.} showed that the broadband electrostatic turbulence in the auroral ionosphere can be identified as the ion cyclotron waves excited by density inhomogeneities and non-uniform electric fields. \cite{Ilyasov2015}
Vranjes and Poedts found that the drift modes in solar corona can accelerate the plasma particles and result in stochastical heating. \cite{Vranjes2009,Vranjes2009a}
Lavorenti \textit{et al.} used the lower-hybrid drift instability to explain the electron energization at Mercury's Magnetopause. \cite{Lavorenti2021}
Besides, the drift waves were also applied to the studies of turbulent transports in fusion plasmas. \cite{Futatani2010,Nishizawa2018}

Numerous observations indicate that space plasmas are always in a non-equilibrium state, and their distributions significantly deviate from the Maxwellian one.
The Kappa distribution is a widely used model to describe the suprathermal particles in different plasma systems, such as solar winds, \citep{Maksimovic1997a,Lazar2017a,LynnB.Wilson2019} discrete auroral arcs, \citep{Ogasawara2017}  the planetary magnetosphere, \citep{Schippers2008,Hapgood2011,Eyelade2021} and cometary plasmas. \citep{Broiles2016,Myllys2019} 
A series of studies reveal that the waves and instabilities in Kappa-distributed plasmas exhibit very different behaviors from those in thermal plasmas. \cite{Gaelzer2014,Gaelzer2016,Guo2021a,Guo2022a,Lazar2023}

The drift wave theory has also been generalized with the assumption of Kappa-distributed particles.
Basu studied the low-frequency drift waves in inhomogeneous magnetized plasmas modeled by anisotropic Kappa distribution. \cite{Basu2008}
The author derived the series representation of linear dispersion relation and found that the drift instability and the current-driven ion-acoustic instability are reduced in comparison with those in the Maxwellian plasmas. \cite{Basu2008}
Huang \textit{et al.} explored the lower-hybrid drift instabilities in the current sheet with Kappa-distributed plasmas and pointed out that the growth rate decreases, but the electromagnetic component increases with enhanced suprathermal particles. \cite{Huang2009}
Batool \textit{et al.} investigated the drift instabilities driven by the ion temperature gradient in non-thermal plasmas and showed that the growth rates are modified by different electron distributions. \cite{Batool2012}

However, in some of the above works, the electrons and ions were assumed to follow the Kappa distribution with the same kappa parameter (e.g., Refs. \onlinecite{Basu2008,Huang2009}), which may not coincide with the recent observations \cite{Eyelade2021}.
In addition, a number of theoretical works suggested that the kappa parameters of different species play different roles in affecting the characteristics of plasmas. \cite{Lazar2023,Guo2021a,Guo2022a}
This paper aims to study the suprathermal effects of different components on drift instabilities in non-uniform magnetized plasmas.

As is known, the dispersion relation in Kappa-distributed magnetized plasmas has a very complicated mathematical form.
In the literature, this difficulty is treated by a few methods, i.e., using the series representation, \cite{Basu2008} defining a new plasma gyroradius function, \cite{Gaelzer2014,Gaelzer2016,Lopez2021} or making some approximations. \cite{Yoon2023}
Both the derivations and results of these treatments are quite complex.
However, in this work, we propose a novel approach to derive the dispersion relation by using an integral representation of Kappa distribution from the superstatistics.
Such an integral representation expresses the Kappa distribution as a superposition of several Maxwellian distributions with different temperatures. 
Based on this method, we obtain a new form of the drift-mode dispersion relation, which has a clear physical meaning and is easy to solve analytically and numerically.
Moreover, this approach can be easily applied to other problems related to Kappa distributions.

The paper is organized as follows.
In Sec. \ref{sec:gdr}, we derive the linear dispersion relation in Kappa-distributed plasmas with the aid of the superstatistics formulas under the local approximations and weak inhomogeneity assumptions.
In Sec. \ref{sec:drift-inst}, the wave frequency and growth rate are obtained analytically and studied numerically.
In addition, we discuss the physical explanations of the suprathermal effects on drift instabilities.
Finally, the conclusions are made in Sec. \ref{sec:sum}.

\section{Linear dispersion relations}
\label{sec:gdr}
\subsection{Model and general dispersion relation}
We consider an inhomogeneous plasma consisting of electrons and ions in a constant magnetic field parallel to the $z$-axis, i.e., $\vb{B} = B \vb{e}_z$.
The inhomogeneities of density and temperature are assumed to be weak and only exist in the $x$-direction.
The plasmas are supposed to be quasi-neutrality, i.e., $n_e \approx n_i$.
In such a system, the constants of motion are the kinetic energy $\varepsilon_\sigma = m_\sigma v^2 /2$ and $X_\sigma = x + v_y / \Omega_\sigma$, where $\Omega_\sigma = q_\sigma B/m_\sigma$ is the cyclotron frequency of ions and electrons for the subscripts $\sigma=i,e$.
The most general form of stationary distribution $f_\sigma^{(0)}$ should be
\begin{equation}
    f_\sigma^{(0)} = f_\sigma^{(0)}(X_\sigma,\varepsilon_\sigma).
\end{equation}
In the assumption of weak inhomogeneities, $f_\sigma^{(0)}$ could be expanded in a series at $X_\sigma = 0$,
\begin{equation}
    f_\sigma^{(0)}(X_\sigma,\varepsilon_\sigma) = \left(1+\frac{X_\sigma}{L_\sigma}\right) F_\sigma(\varepsilon_\sigma),
    \label{eq:f0_series}
\end{equation}
where $L_\sigma$ is the characteristic length of inhomogeneity,
\begin{equation}
    \frac{1}{L_\sigma} = \frac{1}{F_\sigma} \eval{\pdv{f_\sigma^{(0)}}{X_\sigma}}_{X_\sigma=0},
\end{equation}
and $F_\sigma(\varepsilon_\sigma) = f_\sigma^{(0)}(X_\sigma=0,\varepsilon_\sigma)$ is the steady-state distribution in the absence of inhomogeneities.
In such plasmas, we consider a small electrostatic perturbation in the $y$-$z$ plane, leading to the wave vector $\vb{k} = k_y \vb{e}_y + k_z \vb{e}_z$.
By solving the Vlasov-Poisson equations according to the standard method, namely the integration along the unperturbed orbits of particles,
one can derive the general form of linear electrostatic dispersion relation, \cite{Basu2008}
\begin{multline}
    1+\sum_\sigma \frac{q_\sigma^2}{k^2 \epsilon_0 m_\sigma} \int \dd{\vb{v}} 
    \Bigg\{ 
        -2\pdv{}{v_\perp^2} \\
        +\sum_{n=-\infty}^{+\infty} \frac{J^2_n\left(k_y v_\perp/\Omega_\sigma\right)}{\omega-k_z v_z + n\Omega_\sigma} 
        \left[
            k_z \pdv{}{v_z} +2 (\omega - k_z v_z) \pdv{}{v^2_\perp}
            +\frac{k_y}{\Omega_\sigma} \frac{1}{L_\sigma} 
        \right]  
    \Bigg\} 
    F_\sigma(\varepsilon_\sigma) = 0,
    \label{eq:gdr}
\end{multline}
where $v_\perp = \sqrt{v_x^2+v_y^2}$ is the particle speed perpendicular to the magnetic field, $J_n$ is the Bessel function of the first kind, and $F_\sigma(\varepsilon_\sigma)$ is an arbitrary distribution without inhomogeneities.
It is worth noting that the local approximations, $k_y \gg 1/L_\sigma$ and $\rho_\sigma/L_\sigma \ll 1$, are assumed in the derivation of Eq. \eqref{eq:gdr}. \cite{Basu2008,Krall1973}
Here, $\rho_\sigma$ is the Larmor radius for the $\sigma$-species.
In these two approximations,
the former $k_y \gg 1/L_\sigma$ assumes that the characteristic length of inhomogeneity is much larger than the wavelength, indicating that the drift speed is nearly constant in a local region over many wavelengths.
The latter $\rho_\sigma/L_\sigma \ll 1$ denotes the weak inhomogeneity.
More details about the physical explanations of local approximations can be found in the classic textbook \onlinecite{Krall1973}.

\subsection{Kappa distribution for the steady state}
We suppose the stationary distribution without the inhomogeneities is an isotropic Kappa distribution,
\begin{equation}
    F_\sigma = \frac{n_\sigma}{(2\pi \kappa_\sigma \theta_\sigma^2)^{\frac{3}{2}}}
    \frac{\Gamma(\kappa_\sigma+1)}{\Gamma(\kappa_\sigma-1/2)} 
    \left(1+\frac{v^2}{2 \kappa_\sigma \theta^2_\sigma}\right)^{-\kappa_\sigma-1},
    \label{eq:F}
\end{equation}
where $n_\sigma$ is the number density, $\theta_\sigma$ is the thermal speed, and the temperature can be derived,
\begin{equation}
    T_\sigma = \frac{1}{3}\int m_\sigma v^2 \frac{F_\sigma}{n_\sigma} \dd{\vb{v}} = \frac{\kappa_\sigma}{\kappa_\sigma-3/2} m_\sigma \theta_\sigma^2.
    \label{eq:th}
\end{equation}
The kappa parameter $\kappa_\sigma$ has to be in the range $(3/2,+\infty)$ to avoid the divergence of the second-order moment of distribution. \cite{Livadiotis2010a}
In the limit $\kappa_\sigma \rightarrow +\infty$, the distribution \eqref{eq:F} reduces to the Maxwellian one.
In addition, Eq. \eqref{eq:th} shows the relationship between the parameters $\kappa_\sigma$, $T_\sigma$, and $\theta_\sigma$, indicating that only two of them are independent. 
The kappa parameter must be independent of other parameters, so either $T_\sigma$, or $\theta_\sigma$ is $\kappa$-dependent.
In this work, we choose $\theta_\sigma$ and $\kappa_\sigma$ as the independent parameters.
Such a choice could highlight the particle suprathermalization and their consequences. \cite{Lazar2016}

Due to the non-uniform density and temperature, the stationary distribution with the inhomogeneities could be,
\begin{equation}
    f_\sigma^{(0)} = \frac{n_\sigma(X)}{[2\pi \kappa_\sigma \theta_\sigma^2(X)]^{\frac{3}{2}}}
    \frac{\Gamma(\kappa_\sigma+1)}{\Gamma(\kappa_\sigma-1/2)} 
    \left[1+\frac{v^2}{2 \kappa_\sigma \theta^2_\sigma(X)}\right]^{-\kappa_\sigma-1}.
    \label{eq:f0}
\end{equation}

\subsection{Difficulties in deriving the dispersion relation}
\label{sec:diff-dr}
For convenience, we rewrite the dispersion relation \eqref{eq:gdr} as,
\begin{equation}
    1+\sum_\sigma \frac{q_\sigma^2}{k^2 \epsilon_0 m_\sigma} \int \dd{\vb{v}} ( I_1 + I_2 + I_3 ) = 0,
    \label{eq:gdr-split}
\end{equation}
where
\begin{align}
    I_1 &= -\int \dd{\vb{v}} 2\pdv{}{v_\perp^2} F_\sigma(\varepsilon_\sigma), \label{eq:I1} \\
    I_2 &= \int \dd{\vb{v}} \sum_{n=-\infty}^{+\infty} \frac{J^2_n\left(k_y v_\perp/\Omega_\sigma\right)}{\omega-k_z v_z - n\Omega_\sigma} 
        \left[
            k_z \pdv{}{v_z} +2 (\omega - k_z v_z) \pdv{}{v^2_\perp}
        \right]F_\sigma(\varepsilon_\sigma), \label{eq:I2} \\  
    I_3 &= \int \dd{\vb{v}} \sum_{n=-\infty}^{+\infty} \frac{J^2_n\left(k_y v_\perp/\Omega_\sigma\right)}{\omega-k_z v_z - n\Omega_\sigma} \frac{k_y}{\Omega_\sigma} \frac{1}{L_\sigma}F_\sigma(\varepsilon_\sigma). 
         \label{eq:I3}
\end{align}
In Eqs. \eqref{eq:I2} and \eqref{eq:I3}, we have changed $n\rightarrow -n$ in the summation and used the connection formula of the Bessel function, \cite{Olver2010} $J_{-n}(x) = (-1)^n J_n(x)$. 
Before substituting the distribution $F_\sigma$ into these integrals, we need to deal with $I_3$.
In this integral \eqref{eq:I3}, we note
\begin{equation}
    \frac{F_\sigma}{L_\sigma} = \left. \pdv{f_\sigma^{(0)}}{X_\sigma} \right|_{X_\sigma=0}
                              = \left(
                                \left. \dv{\ln n_\sigma}{X_\sigma} \right|_ {X_\sigma=0}
                                + \left. \dv{T_\sigma}{X_\sigma} \right|_{X_\sigma=0} \pdv{}{T_\sigma} 
                              \right) F_\sigma,
    \label{eq:F2L}
\end{equation}
by inserting Eq. \eqref{eq:f0}.
Here, we suppress the arguments of functions for concise.
Because the variable $x$ appearing in $f_\sigma^{(0)}$ \eqref{eq:f0} is only in the combination $x+v_y/\Omega_\sigma$, one can rewrite the above equation \eqref{eq:F2L} as,
\begin{equation}
    \frac{F_\sigma}{L_\sigma} = \left( \left. \dv{\ln n_\sigma}{x} \right|_ {x=0} + \left. \dv{T_\sigma}{x} \right|_{x=0} \pdv{}{T_\sigma} \right) F_\sigma,
    \label{eq:F2Lre}
\end{equation}
If we define a drift frequency operator,
\begin{equation}
    \hat{\omega}_d = \frac{k_y T_\sigma}{\Omega_\sigma m_\sigma} \left( \left. \dv{\ln n_\sigma}{x} \right|_ {x=0} + \left. \dv{T_\sigma}{x} \right|_{x=0} \pdv{}{T_\sigma} \right),
    \label{eq:wd}
\end{equation}
then
\begin{equation}
    \frac{F_\sigma}{L_\sigma} = \frac{\Omega_\sigma m_\sigma}{k_y T_\sigma}\hat{\omega}_d F_\sigma.
\end{equation}
Because the operator $\hat{\omega}_d$, including the partial derivative with respect to the temperature $\pdv*{}{T_\sigma}$, only acts on $F_\sigma(\varepsilon_\sigma)$ in the integrand of $I_3$, one can place $\hat{\omega}_d$ outside the integral,
\begin{equation}
    I_3 = \frac{m_\sigma}{T_\sigma}\hat{\omega}_d \int \dd{\vb{v}} \sum_{n=-\infty}^{+\infty} \frac{J^2_n\left(k_y v_\perp/\Omega_\sigma\right)}{\omega-k_z v_z - n\Omega_\sigma} F_\sigma(\varepsilon_\sigma). 
    \label{eq:I3-wd}
\end{equation}
Now, we substitute $F_\sigma(\varepsilon_\sigma)$ \eqref{eq:F} into Eqs. \eqref{eq:I1}, \eqref{eq:I2}, and \eqref{eq:I3-wd}.
$I_1$ could be directly integrated,
\begin{equation}
    I_1 = n_\sigma \frac{m_\sigma}{T_\sigma}\frac{\kappa_\sigma-1/2}{\kappa_\sigma-3/2}.
    \label{eq:I1-solution}
\end{equation}
By interchanging the sum and integral, $I_2$ and $I_3$ turn out to be,
\begin{multline}
    I_2 = -\frac{m_\sigma}{T_\sigma} \omega 
    \frac{n_\sigma}{(2\pi \kappa_\sigma \theta_\sigma^2)^{\frac{3}{2}}} 
    \frac{\Gamma(\kappa_\sigma+1)}{\Gamma(\kappa_\sigma-1/2)} 
    \frac{\kappa_\sigma+1}{\kappa_\sigma-3/2} \\
    \times
    \sum_{n=-\infty}^{+\infty} \int_0^{+\infty} \dd{v_\perp} \int_{-\infty}^{+\infty} \dd{v_z}
    \frac{2\pi v_\perp J^2_n\left(k_y v_\perp/\Omega_\sigma\right)}{\omega-k_z v_z - n\Omega_\sigma} 
    \left(1+\frac{v^2}{2 \kappa_\sigma \theta^2_\sigma}\right)^{-\kappa_\sigma-2},
    \label{eq:I2-int}
\end{multline}
and
\begin{multline}
    I_3 = \frac{m_\sigma}{T_\sigma} \hat{\omega}_d
    \frac{n_\sigma}{(2\pi \kappa_\sigma \theta_\sigma^2)^{\frac{3}{2}}} 
    \frac{\Gamma(\kappa_\sigma+1)}{\Gamma(\kappa_\sigma-1/2)} \\
    \times
    \sum_{n=-\infty}^{+\infty} \int_0^{+\infty} \dd{v_\perp} \int_{-\infty}^{+\infty} \dd{v_z}
    \frac{2\pi v_\perp J^2_n\left(k_y v_\perp/\Omega_\sigma\right)}{\omega-k_z v_z - n\Omega_\sigma} 
    \left(1+\frac{v^2}{2 \kappa_\sigma \theta^2_\sigma}\right)^{-\kappa_\sigma-1}.
    \label{eq:I3-int}
\end{multline}

In the Maxwellian limit, the last terms of the integrands in Eqs. \eqref{eq:I2-int} and \eqref{eq:I3-int} tend to,
\begin{align}
    &\lim_{\kappa_\sigma \rightarrow +\infty} 
    \left(1+\frac{v^2}{2 \kappa_\sigma \theta^2_\sigma}\right)^{-\kappa_\sigma-2} \notag\\
    =&\lim_{\kappa_\sigma \rightarrow +\infty} 
    \left(1+\frac{v^2}{2 \kappa_\sigma \theta^2_\sigma}\right)^{-\kappa_\sigma-1} \notag\\
    =&\exp(-\frac{v^2}{2 \theta^2_\sigma})\notag \\
    =&\exp(-\frac{v_\perp^2}{2 \theta^2_\sigma}) \exp(-\frac{v_z^2}{2 \theta^2_\sigma}),
\end{align}
so the integrals of $\int \dd{v_\perp}$ and $\int \dd{v_z}$ can be separately calculated in Eqs. \eqref{eq:I2-int} and \eqref{eq:I3-int} when $\kappa_\sigma \rightarrow +\infty$.
However, for the finite kappa values, the integrations over ${v_\perp}$ and ${v_z}$ in Eqs. \eqref{eq:I2-int} and \eqref{eq:I3-int} are coupled, so it is difficult to derive the analytical solution of $I_2$ and $I_3$.
In the literature, the methods to analyze the integrals similar to \eqref{eq:I2-int} and \eqref{eq:I3-int} are to expand them into series, \cite{Basu2008} 
calculate the integral over $v_z$ and define the remaining integral over $v_\perp$ as a new function, \cite{Gaelzer2014,Gaelzer2016} make some appropriate approximations, \cite{Yoon2023} or carry out numerical calculations. \cite{Lopez2021} 
In this study, we propose a new approach by adopting the formulas from the superstatistics to simplify $I_2$ and $I_3$, showing lots of advantages in the following content.

\subsection{Superstatistics formulas}
The superstatistics theory interprets the nonextensive distribution as a superposition of many Maxwellian distributions with fluctuating inverse temperatures, \cite{Beck2001a,Beck2003}
\begin{equation}
    [1+(q-1)\beta_0 E]^{-\frac{1}{q-1}} = \int_0^{+\infty} g(\beta) e^{-\beta E} \dd{\beta},
    \label{eq:sps-origin}
\end{equation}
where $g(\beta)$ is the gamma distribution,
\begin{equation}
    g(\beta) = \frac{(q-1)^{-\frac{1}{q-1}}}{\beta_0 \Gamma[1/(q-1)]} \left(\frac{\beta}{\beta_0}\right)^{\frac{1}{q-1}-1} \exp[-\frac{\beta}{(q-1)\beta_0}].
    \label{eq:g}
\end{equation}
Due to the connection and the similarity between the nonextensive distribution and the Kappa one, \cite{Livadiotis2009,Livadiotis2015a} 
we can rewrite the above expressions \eqref{eq:sps-origin} and \eqref{eq:g} into two useful forms related to the Kappa distribution.
With a group of transformations,
\begin{equation}
    \label{eq:trans-g1} 
    E = \frac{v^2}{2 \theta_\sigma^2}, \quad 
    (q-1)\beta_0 = \frac{1}{\kappa_\sigma}, \quad 
    \frac{1}{q-1} = \kappa_\sigma+2, 
\end{equation}
Eq. \eqref{eq:sps-origin} yields,
\begin{equation}
    \left(1+\frac{v^2}{2 \kappa_\sigma \theta^2_\sigma}\right)^{-\kappa_\sigma-2} 
    = \int_0^{+\infty} g_1(b_\sigma) \exp\left(-b_\sigma \frac{v^2}{2 \theta^2_\sigma}\right) \dd{b_\sigma},
    \label{eq:sup-int-g1}
\end{equation}
where 
\begin{equation}
    g_1(b_\sigma) = \frac{\kappa_\sigma^{\kappa_\sigma+2}b_\sigma^{\kappa_\sigma+1}}{\Gamma(\kappa_\sigma+2)} e^{-\kappa_\sigma b_\sigma}.
    \label{eq:g1}
\end{equation}
and we have changed the symbol $\beta$ to $b_\sigma$ to avoid confusion with the plasma beta.
With another group of transformations,
\begin{equation}
    \label{eq:trans-g2} 
    E = \frac{v^2}{2 \theta_\sigma^2}, \quad 
    (q-1)\beta_0 = \frac{1}{\kappa_\sigma}, \quad 
    \frac{1}{q-1} = \kappa_\sigma+1, 
\end{equation}
one obtains,
\begin{equation}
    \left(1+\frac{v^2}{2 \kappa_\sigma \theta^2_\sigma}\right)^{-\kappa_\sigma-1} 
    = \int_0^{+\infty} g_2(b_\sigma) \exp\left(-b_\sigma \frac{v^2}{2 \theta^2_\sigma}\right) \dd{b_\sigma},
    \label{eq:sup-int-g2}
\end{equation}
where 
\begin{equation}
    g_2(b_\sigma) = \frac{\kappa_\sigma^{\kappa_\sigma+1}b_\sigma^{\kappa_\sigma}}{\Gamma(\kappa_\sigma+1)} e^{-\kappa_\sigma b_\sigma}.
    \label{eq:g2}
\end{equation}

\subsection{Dispersion relation in Kappa-distributed plasmas with inhomogeneous density and temperature}
\label{sec:dr-kappa}
Taking Eq. \eqref{eq:sup-int-g1} back into $I_2$ \eqref{eq:I2-int} and interchanging the order of integral, we have,
\begin{multline}
    I_2 = -\frac{m_\sigma}{T_\sigma} \omega 
    \frac{n_\sigma}{(2\pi \kappa_\sigma \theta_\sigma^2)^{\frac{3}{2}}} 
    \frac{\Gamma(\kappa_\sigma+1)}{\Gamma(\kappa_\sigma-1/2)} 
    \frac{\kappa_\sigma+1}{\kappa_\sigma-3/2}
    \sum_{n=-\infty}^{+\infty} \int_0^{+\infty} \dd{b_\sigma} g_1(b_\sigma)\\
    \times
    \int_0^{+\infty} \dd{v_\perp} 
    2\pi v_\perp J^2_n\left(\frac{k_y v_\perp}{\Omega_\sigma}\right)\exp\left(-b_\sigma \frac{v_\perp^2}{2 \theta^2_\sigma}\right)
    \int_{-\infty}^{+\infty} \dd{v_z}
    \frac{\exp\left(-b_\sigma \frac{v_z^2}{2 \theta^2_\sigma}\right)}{\omega-k_z v_z - n\Omega_\sigma}.
    \label{eq:I2-dev1}
\end{multline}
The integrations over $v_\perp$ and $v_z$ are well known in the traditional kinetic theory of Maxwellian plasmas, and the solutions are, \cite{Weiland2000, Ichimaru2004}
\begin{equation}
    \int_0^{+\infty} \dd{v_\perp} 
    2\pi v_\perp J^2_n\left(\frac{k_y v_\perp}{\Omega_\sigma}\right)\exp\left(-b_\sigma \frac{v_\perp^2}{2 \theta^2_\sigma}\right)
    = \frac{2\pi\theta_\sigma^2}{b_\sigma}\Lambda_n \left(\frac{k_y^2 \theta_\sigma^2}{\Omega_\sigma^2 b_\sigma}\right),
    \label{eq:int_vperp}
\end{equation}
and
\begin{equation}
    \int_{-\infty}^{+\infty} \dd{v_z}
    \frac{\exp\left(-b_\sigma \frac{v_z^2}{2 \theta^2_\sigma}\right)}{\omega-k_z v_z - n\Omega_\sigma}
    = - \frac{\sqrt{2\pi}}{w-n\Omega_\sigma}\frac{\theta_\sigma}{\sqrt{b_\sigma}}
        \left[W\left(\frac{\omega-n\Omega_\sigma}{k_z\theta_\sigma/\sqrt{b_\sigma}}\right)-1\right],
    \label{eq:int_vz}
\end{equation}
where $\Lambda_n(x) =\exp(-x) I_n(x)$ is defined with the modified Bessel function $I_n(x)$,
and $W(z)$ is the plasma dispersion $W$-function defined by, \cite{Weiland2000}
\begin{equation}
    W(z) = \frac{1}{\sqrt{2\pi}} \int_{-\infty}^{+\infty}\frac{x}{x-z}\exp\left(-\frac{x^2}{2}\right) \dd{x}.
    \label{eq:disperfunc-w}
\end{equation}
After rearrangements, $I_2$ becomes,
\begin{equation}
    I_2 = \frac{m_\sigma}{T_\sigma} n_\sigma \omega \frac{\kappa_\sigma-1/2}{\kappa_\sigma-3/2} 
    \sum_{n=-\infty}^{+\infty} \int_0^{+\infty} \dd{b_\sigma} 
    \frac{G_1(b_\sigma)}{\omega-n\Omega_\sigma}
    \left[ W\left(\frac{\omega-n\Omega_\sigma}{k_z\theta_\sigma/\sqrt{b_\sigma}}\right) -1 \right]
    \Lambda_n \left(\frac{k_y^2 \theta_\sigma^2}{\Omega_\sigma^2 b_\sigma}\right),
    \label{eq:I2-solution}
\end{equation}
where $G_1(b_\sigma)$ is given by,
\begin{equation}
    G_1(b_\sigma) = \frac{\kappa_\sigma^{\kappa_\sigma+1/2}b_\sigma^{\kappa_\sigma-1/2}}{\Gamma(\kappa_\sigma+1/2)} e^{-\kappa_\sigma b_\sigma}.
    \label{eq:G1}
\end{equation}
By similar procedures, $I_3$ is derived,
\begin{equation}
    I_3 = -\frac{m_\sigma}{T_\sigma} n_\sigma \hat{\omega}_d \frac{\kappa_\sigma-1/2}{\kappa_\sigma-3/2} 
    \sum_{n=-\infty}^{+\infty} \int_0^{+\infty} \dd{b_\sigma} 
    \frac{G_2(b_\sigma)}{\omega-n\Omega_\sigma}
    \left[ W\left(\frac{\omega-n\Omega_\sigma}{k_z\theta_\sigma/\sqrt{b_\sigma}}\right) -1 \right]
    \Lambda_n \left(\frac{k_y^2 \theta_\sigma^2}{\Omega_\sigma^2 b_\sigma}\right),
    \label{eq:I3-solution}
\end{equation}
where $G_2(b_\sigma)$ is,
\begin{equation}
    G_2(b_\sigma) = \left(\kappa_\sigma-\frac{3}{2}\right)\frac{\kappa_\sigma^{\kappa_\sigma-1/2}b_\sigma^{\kappa_\sigma-3/2}}{\Gamma(\kappa_\sigma+1/2)} e^{-\kappa_\sigma b_\sigma}.
    \label{eq:G2}
\end{equation}
We stress that $G_1(b_\sigma)$ \eqref{eq:G1} and $G_2(b_\sigma)$ \eqref{eq:G2} are the gamma distributions with different parameters,
and their properties are demonstrated in Appendix \ref{sec:g1g2}.
The integrals over $b_\sigma$ in $I_2$ \eqref{eq:I2-solution} and $I_3$ \eqref{eq:I3-solution} can be regarded as a superposition of,
\begin{equation}
    \frac{1}{\omega-n\Omega_\sigma}
    \left[ W\left(\frac{\omega-n\Omega_\sigma}{k_z\theta_\sigma/\sqrt{b_\sigma}}\right) -1 \right]
    \Lambda_n \left(\frac{k_y^2 \theta_\sigma^2}{\Omega_\sigma^2 b_\sigma}\right),
    \label{eq:int-kernel}
\end{equation}
with different weight functions $G_1(b_\sigma)$ and $G_2(b_\sigma)$, respectively.
Finally, substituting Eqs. \eqref{eq:I1-solution}, \eqref{eq:I2-solution} and \eqref{eq:I3-solution} into Eq. \eqref{eq:gdr-split}, we find the linear dispersion relation for inhomogeneous Kappa-distributed plasmas,
\begin{multline}
    1+\sum_\sigma \frac{1}{k^2 \lambda_{\kappa\sigma}^2} \Bigg\{ 1+
    \sum_{n=-\infty}^{+\infty} \int_0^{+\infty} \dd{b_\sigma} \\
    \times
    \frac{G_1(b_\sigma)\omega - G_2(b_\sigma) \hat{\omega}_d}{\omega-n\Omega_\sigma}
    \left[
    W\left(\frac{\omega-n\Omega_\sigma}{k_z\theta_\sigma/\sqrt{b_\sigma}}\right) -1
    \right]
    \Lambda_n \left(\frac{k_y^2 \theta_\sigma^2}{\Omega_\sigma^2 b_\sigma}\right)
    \Bigg\}=0,
    \label{eq:dr-kappa-w}
\end{multline}
where $\lambda_{\kappa\sigma}$ is the Debye length of Kappa-distributed plasmas, \cite{Lazar2022a}
\begin{equation}
   \lambda_{\kappa\sigma} = \sqrt{\frac{\kappa_\sigma}{\kappa_\sigma-1/2}}
   \sqrt{\frac{\epsilon_0 m_\sigma \theta_\sigma^2}{n_\sigma q_\sigma^2}}.
   \label{eq:l_D}
\end{equation}
In the limit $\kappa_\sigma \rightarrow +\infty$, both $G_1(b_\sigma)$ and $G_2(b_\sigma)$ are reduced to the delta function $\delta(b_\sigma-1)$ (see proof in Appendix \ref{sec:g1g2}), restoring Eq. \eqref{eq:dr-kappa-w} to the dispersion relation in Maxwellian plasmas, \cite{Ichimaru2004}
\begin{equation}
    1+\sum_\sigma \frac{1}{k^2 \lambda_{D\sigma}^2} \left\{ 1+
    \sum_{n=-\infty}^{+\infty}
    \frac{\omega - \hat{\omega}_d}{\omega-n\Omega_\sigma}
    \left[
    W\left(\frac{\omega-n\Omega_\sigma}{k_z\theta_\sigma}\right) -1
    \right]
    \Lambda_n \left(\frac{k_y^2 \theta_\sigma^2}{\Omega_\sigma^2}\right)
    \right\}=0.
    \label{eq:dr-maxwellian}
\end{equation}
Eq. \eqref{eq:dr-kappa-w} indicates that the dispersion relation in Kappa-distributed plasmas could be understood as a superposition of Maxwellian dispersion relations with different thermal speeds $\theta_\sigma/\sqrt{b_\sigma}$ (due to the varied $b_\sigma$). 

In the integral of the dispersion relation \eqref{eq:dr-kappa-w}, different terms are weighted by different functions $G_1(b_\sigma)$ and $G_2(b_\sigma)$.
The reason is that the operator $\hat{\omega}_d$ \eqref{eq:wd} contains a factor of temperature $T_\sigma$ rather than thermal speed $\theta_\sigma$.
If we define a modified operator,
\begin{equation}
    \hat{\omega}_{d}'(\theta_\sigma) = \frac{k_y \theta^2_\sigma}{\Omega_\sigma} \left( \left. \dv{\ln n_\sigma}{x} \right|_ {x=0} + \left. \dv{T_\sigma}{x} \right|_{x=0} \pdv{}{T_\sigma} \right) = \frac{\kappa_\sigma-3/2}{\kappa_\sigma} \hat{\omega}_d,
    \label{eq:wd-modified}
\end{equation}
the dispersion relation \eqref{eq:dr-kappa-w} can be equivalently rewritten as,
\begin{multline}
    1+\sum_\sigma \frac{1}{k^2 \lambda_{\kappa\sigma}^2} \Bigg\{ 1+
    \sum_{n=-\infty}^{+\infty} \int_0^{+\infty} \dd{b_\sigma} G_1(b_\sigma) \\
    \times 
    \frac{\omega - \hat{\omega}_{d}'(\theta/\sqrt{b_\sigma})}{\omega-n\Omega_\sigma}
    \left[
    W\left(\frac{\omega-n\Omega_\sigma}{k_z\theta_\sigma/\sqrt{b_\sigma}}\right) -1
    \right]
    \Lambda_n \left(\frac{k_y^2 \theta_\sigma^2}{\Omega_\sigma^2 b_\sigma}\right)
    \Bigg\}=0.
    \label{eq:dr-kappa-w-modified}
\end{multline}
However, the operator $\hat{\omega}_d$ has a more clear physical meaning than the modified one $\hat{\omega}_{d}'$.
For instance, if there is no temperature gradient $\dv*{T_\sigma}{x}=0$, then $\hat{\omega}_d$ regresses to the drift frequency ${\omega}_d = k_y u_\sigma$, where $u_\sigma$ is the drift speed of $\sigma$-particle due to the inhomogeneous density.
As we know, the drift frequency $\omega_d$ is the result of the fluid theory so it is independent of the particle distribution. \cite{Ichimaru2004}
Therefore, we still adopt Eq. \eqref{eq:dr-kappa-w} as the preferred expression of dispersion relation in the present paper.

\section{Drift mode instabilities}
\label{sec:drift-inst}
In this section, 
we focus on the low-frequency drift instabilities driven by the inhomogeneous density in Kappa-distributed plasmas.
Here, the temperature is presumed to be uniform.
As a result, the operator $\hat{\omega}_d$
\begin{equation}
    \hat{\omega}_d = \omega_{N\sigma} = \frac{k_y T_\sigma}{\Omega_\sigma m_\sigma} \left. \dv{\ln n_\sigma}{x} \right|_ {x=0},
    \label{eq:drift-freq}
\end{equation}
does not include the partial derivative operator $\pdv*{}{T_\sigma}$ any longer.
Because of the quasi-neutrality, it is apparent that,
\begin{equation}
    \frac{\omega_{Ni}}{\omega_{Ne}} = -\frac{T_i}{T_e}.
    \label{eq:omegai2e}
\end{equation}
Additionally, in the low-frequency condition $\omega \ll \Omega_i$, we can only retain the $n=0$ term in the sum of the dispersion relation \eqref{eq:dr-kappa-w}, i.e.,
\begin{multline}
    1+\sum_\sigma \frac{1}{k^2 \lambda_{\kappa\sigma}^2} \Bigg\{ 1+
    \int_0^{+\infty} \dd{b_\sigma} 
    \left[G_1(b_\sigma) - G_2(b_\sigma) \frac{\omega_{N\sigma}}{\omega}\right] \\
    \times
    \left[
    W\left(\frac{\omega}{k_z\theta_\sigma/\sqrt{b_\sigma}}\right) -1
    \right]
    \Lambda_0 \left(\frac{k_y^2 \rho_\sigma^2}{b_\sigma}\right)
    \Bigg\}=0,
    \label{eq:dr-kappa-w-lf}
\end{multline}
where $\rho_\sigma = \theta_\sigma/|\Omega_\sigma|$ is the average Larmor radius for $\sigma$-particles.

\subsection{Approximate solutions in the regime \texorpdfstring{$\theta_i\ll\omega/k_z\ll\theta_e$}{θi<<ω/kz<<θe}}
\label{sec:appsol}
Expanding the dispersion relation \eqref{eq:dr-kappa-w-lf} with $\theta_i\ll\omega/k_z\ll\theta_e$, we attain the dielectric function (see details in Appendix \ref{sec:expansion}),
\begin{multline}
    \epsilon = 
    1+\frac{1}{k^2 \lambda_{\kappa e}^2}+\frac{1}{k^2 \lambda_{\kappa i}^2}\left(1+\frac{\omega_{Ni}}{\omega}A_2-A_1\right) \\
    + i \sqrt{\frac{\pi}{2}} \left[
    \frac{\xi_e}{k^2 \lambda_{\kappa e}^2}
    \frac{\Gamma(\kappa_e)\sqrt{\kappa_e}}{\Gamma(\kappa_e+1/2)}
    \left(1-\frac{\kappa_e-3/2}{\kappa_e}\frac{\omega_{Ne}}{\omega}\right)
    +\frac{\xi_i}{k^2 \lambda_{\kappa i}^2}
    \left(B_1-B_2\frac{\omega_{Ni}}{\omega}\right)
        \right],
    \label{eq:epsilon}
\end{multline}
where $\xi_{e,i} = \omega/(k_z \theta_{e,i})$, and the notations $A_1$, $A_2$, $B_1$, and $B_2$ are defined by,
\begin{align}
    & A_1 = \int_0^{+\infty} \dd{b_i} 
    G_1(b_i) \Lambda_0 \left(\frac{k_y^2 \rho_i^2}{b_i}\right),\label{eq:A1} \\
    & A_2 = \int_0^{+\infty} \dd{b_i} 
    G_2(b_i) \Lambda_0 \left(\frac{k_y^2 \rho_i^2}{b_i}\right),\label{eq:A2} \\
    & B_1 = \int_0^{+\infty} \dd{b_i} 
    G_1(b_i) \sqrt{b_i} e^{-\frac{\xi_i^2 b_i}{2}}
    \Lambda_0 \left(\frac{k_y^2 \rho_i^2}{b_i}\right),\label{eq:B1} \\
    & B_2 = \int_0^{+\infty} \dd{b_i} 
    G_2(b_i) \sqrt{b_i} e^{-\frac{\xi_i^2 b_i}{2}}
    \Lambda_0 \left(\frac{k_y^2 \rho_i^2}{b_i}\right).\label{eq:B2}
\end{align}
We let $\omega = \omega_r + i \gamma$ and presume the growth/damping rate is much less than the real frequency of the drift mode, i.e., $|\gamma| \ll \omega_r$.
Then, $\omega_r$ can be derived by setting $\Re(\epsilon) = 0$,
\begin{equation}
    \omega_r = -\frac{\omega_{Ni} A_2}
            {
                1+k^2\lambda_{\kappa i}^2 + \lambda_{\kappa i}^2/\lambda_{\kappa e}^2 - A_1
            },
    \label{eq:wr}
\end{equation}
and the growth rate $\gamma = -\Im(\epsilon)/[\pdv*{\Re(\epsilon)}{\omega_r}]$ is,
\begin{align}
    \gamma   &= \gamma_e + \gamma_i, \label{eq:gamma} \\
    \gamma_e &= \sqrt{\frac{\pi}{2}} \frac{\omega_r^2}{A_2}
        \frac{1}{k_z \theta_e} 
        \frac{\kappa_e-1/2}{\kappa_e-3/2}\frac{\kappa_i-3/2}{\kappa_i-1/2} 
        \frac{\Gamma(\kappa_e)\sqrt{\kappa_e}}{\Gamma(\kappa_e+1/2)} 
        \left(\frac{\kappa_e-3/2}{\kappa_e}-\frac{\omega_r}{\omega_{Ne}}\right) , 
        \label{eq:gamma_e}\\
    \gamma_i &= -\sqrt{\frac{\pi}{2}} \frac{\omega_r^2}{A_2}
        \frac{1}{k_z \theta_i} \left(B_2-B_1\frac{\omega_r}{\omega_{Ni}}\right), 
        \label{eq:gamma_i}
\end{align}
where $\gamma_e$ and $\gamma_i$ denote the electron and ion contributions, respectively.
It should be noticed that $A_1>0$, $A_2>0$, $B_1>0$, and $B_2>0$ 
because the integrands in Eqs. \eqref{eq:A1}-\eqref{eq:B2} are all positive.
Further, $A_1<1$ is proved because $G_1(b_\sigma)$ is a normalized gamma distribution, and $1>\Lambda_0(x)>0$ for $x>0$. \cite{Olver2010}
Hence, the ion term \eqref{eq:gamma_i} is negative-definite, leading to a damping contribution as the same in the Maxwellian situation.
Due to the exponent functions in $B_1$ \eqref{eq:B1} and $B_2$ \eqref{eq:B2}, with the approximation $\xi_i \gg 1$, the ion Landau damping can be neglected.
However, the characteristic of electron contribution is interesting. 
In the Maxwellian limit, the electron term $\gamma_e$ is positive-definite, which means the drift mode is always unstable, so the drift instability is also called universal instability. \cite{Ichimaru2004}
But in the Kappa case, the condition for $\gamma_e>0$ is,
\begin{equation}
    \frac{\kappa_e-3/2}{\kappa_e}-\frac{\omega_r}{\omega_{Ne}}>0, 
    \label{eq:cond-gamma-dev}
\end{equation}
which is not always fulfilled.
When the ion gyroradius is negligible $k_y^2 \rho_i^2 \ll 1$, $A_1$ and $A_2$ approximate to,
\begin{equation}
    A_1 
        \approx 1-\frac{\kappa_i}{\kappa_i-1/2}k_y^2 \rho_i^2,
    \label{eq:A1-approx}
\end{equation}
and
\begin{equation}
    A_2 
        \approx \frac{\kappa_i-3/2}{\kappa_i-1/2}-\frac{\kappa_i}{\kappa_i-1/2}k_y^2 \rho_i^2,
    \label{eq:A2-approx}
\end{equation}
by expanding the $\Lambda_0$ in Eqs. \eqref{eq:A1} and \eqref{eq:A2}, and calculating the integrals.
After substituting Eqs. \eqref{eq:A1-approx}, \eqref{eq:A2-approx}, and \eqref{eq:omegai2e} into the real frequency \eqref{eq:wr}, one arrives at,
\begin{align}
    \frac{\omega_r}{\omega_{Ne}} 
    &= \frac{1-\frac{\kappa_i}{\kappa_i-3/2}k_y^2\rho_i^2}
    {
        \frac{\kappa_e-1/2}{\kappa_e-3/2}+\frac{T_e}{T_i}\left(k^2\lambda_{\kappa i}^2\frac{\kappa_i-1/2}{\kappa_i-3/2} +\frac{\kappa_i}{\kappa_i-3/2}k_y^2\rho_i^2 \right)
    } \notag \\
    &\approx \frac{1-\frac{\kappa_i}{\kappa_i-3/2}k_y^2\rho_i^2}
    {
        \frac{\kappa_e-1/2}{\kappa_e-3/2}+\frac{m_e\theta_e^2}{m_i\theta_i^2}\frac{\kappa_e}{\kappa_e-3/2}k_y^2\rho_i^2
    },
    \label{eq:omegar2e}
\end{align}
where we use the approximation $k^2 \lambda_{\kappa i}^2 \ll k_y^2 \rho_i^2$ for the following reasons.
First, $\lambda_{\kappa i} \ll \rho_i$ demonstrates that the cyclotron motion is in a quasi-neutral background, implying that the drift motion is generated only by the density gradient.
Such an approximation is valid in many real plasmas. \cite{Ichimaru2004,Malaspina2021}
Second, 
in terms of the fluid theory, \cite{Ichimaru2004}
the drift wave should propagate with a sufficient small $k_z$ to ensure $|\omega_{Ne}/k_z| \gg c_s$ so that
\begin{equation}
    \Big|\frac{k_y}{k_z}\Big| \gg \frac{L_N}{\rho_i} \sqrt{\frac{m_i \theta_i^2}{m_e \theta_e^2}} \frac{\kappa_e-3/2}{\sqrt{\kappa_e}\sqrt{\kappa_e-1/2}},
\end{equation}
where $L_N^{-1}=(\dv*{\ln n_e}{x})_{x=0}=(\dv*{\ln n_i}{x})_{x=0}$ is the characteristic length of inhomogeneous density, and $c_s$ is the ion-acoustic speed defined by \cite{Mace1998} 
$ c_s = \sqrt{\kappa_e/(\kappa_e-1/2)}\sqrt{m_e/m_i}\theta_e $
in Kappa-distributed plasmas.
Because the weak inhomogeneity requires $L_N \gg \rho_i$, and $m_e\theta_e^2$ is usually comparable to $m_i \theta_i^2$, one can neglect $k_z^2 \lambda_{\kappa i}^2$ compared with $k_y^2 \rho_i^2$.
Eventually, 
by taking Eq. \eqref{eq:omegar2e} back into Eq. \eqref{eq:cond-gamma-dev},
the criterion for $\gamma_e>0$ is derived,
\begin{equation}
    k_y^2 \rho_i^2 > K_y^2 = \frac{1}{2\kappa_e}\frac{1}{\kappa_i/(\kappa_i-3/2)+(m_e/m_i)(\theta_e^2/\theta_i^2)},
    \label{eq:kyrhoi-cond-2}
\end{equation}
where $K_y$ is the dimensionless critical/threshold wavenumber scaled by the average ion gyroradius $\rho_i$.
Such an instability condition can be attributed to the suprathermalization of electrons
because $K_y$ vanishes when $\kappa_e \rightarrow +\infty$.
One may argue that $\kappa_i \rightarrow 3/2$ can also lead to zero critical wavenumber $K_y$.
We must stress that the limit $\kappa_{e,i} \rightarrow +\infty$ is available, which represents the Maxwellian equilibrium, 
but the limit $\kappa_{e,i} \rightarrow 3/2$ is unreachable because $\kappa_{e,i} = 3/2$ results in an infinite temperature. \cite{Livadiotis2010a}
Therefore, the suprathermalization of ions cannot cancel but only lower the critical wavenumber $K_y$.

Besides, the derivations \eqref{eq:epsilon}-\eqref{eq:kyrhoi-cond-2} indicate the distinct advantages of the integral representation of the dispersion relation \eqref{eq:dr-kappa-w-lf}.
To solve the analytical real frequency and growth rate, we simply follow the same procedures dealing with the Maxwellian plasmas and only use the expansions of the standard plasma dispersion function.
In the present approach, there is no necessity to introduce a generalized plasma dispersion function modified by the Kappa distribution, \cite{Mace1995,Basu2008} or a suprathermal plasma gyroradius function, \cite{Gaelzer2014,Gaelzer2016} saving the studies on the properties of these newly defined functions.

\subsection{Numerial analysis}
\label{sec:num}
To obtain more accurate results, we numerically calculate the dispersion relation \eqref{eq:dr-kappa-w-lf}.
For convenience, dimensionless parameters are used in the numerical investigation.
The frequency and the growth rate are calculated in the unit of ion cyclotron frequency $\Omega_i$.
The length, as well as the wave vector, is normalized by ion gyroradius $\rho_i$.
We adopt some typical parameters in space plasmas.
The kappa values of electrons were observed to be in the range of $2<\kappa_e<5$ in the solar wind,~\cite{Maksimovic1997a} $3<\kappa_e<6$ in the discrete auroral arcs,~\cite{Ogasawara2017} and $4\le \kappa_e \le 5$ in the magnetosphere.~\cite{Eyelade2021}
For ions, it was reported to be $2.4<\kappa_i<4.7$ in the solar wind~\cite{Collier1996, Chotoo2000} and $5 \le \kappa_i \le 8$ in the magnetosphere.~\cite{Eyelade2021}
Therefore, we choose the kappa parameters in a typical range of $2<\kappa_{e,i}<8$ for both electrons and ions.
The mass ratio between ions and electrons is $m_i/m_e=1836$, and the ratio of thermal speeds is assumed to be $\theta_e/\theta_i = 100$.
Consequently, the electron-to-ion temperature ratio is determined by the kappa parameters, 
\begin{equation}
    \frac{T_e}{T_i} = \frac{m_e \theta_e^2}{m_i \theta_i^2} \frac{\kappa_e}{\kappa_i}\frac{\kappa_i-3/2}{\kappa_e-3/2}
    .
    \label{eq:Te2Ti}
\end{equation}
Considering the kappa values $2<\kappa_{e,i}<8$, one finds the temperature ratio is in the range $1.68<T_e/T_i<17.70$.
The characteristic length of the density inhomogeneity is set as $L_N/\rho_i = 10$, with $L_N^{-1}=(\dv*{\ln n_e}{x})_{x=0}=(\dv*{\ln n_i}{x})_{x=0}$.
The ion oscillation frequency is $\omega_{pi}/\Omega_i = 100$.
These parameter selections are inspired from the observations of space plasmas.~\cite{Guo2022b}
The other physical quantities can be inferred according to the above parameters.
Besides, the integrals in the dispersion relation \eqref{eq:dr-kappa-w-lf} are implemented by generalized Gauss-Laguerre quadrature with $100$ terms.

\begin{figure}
    \centering
    \includegraphics[width=8.5cm]{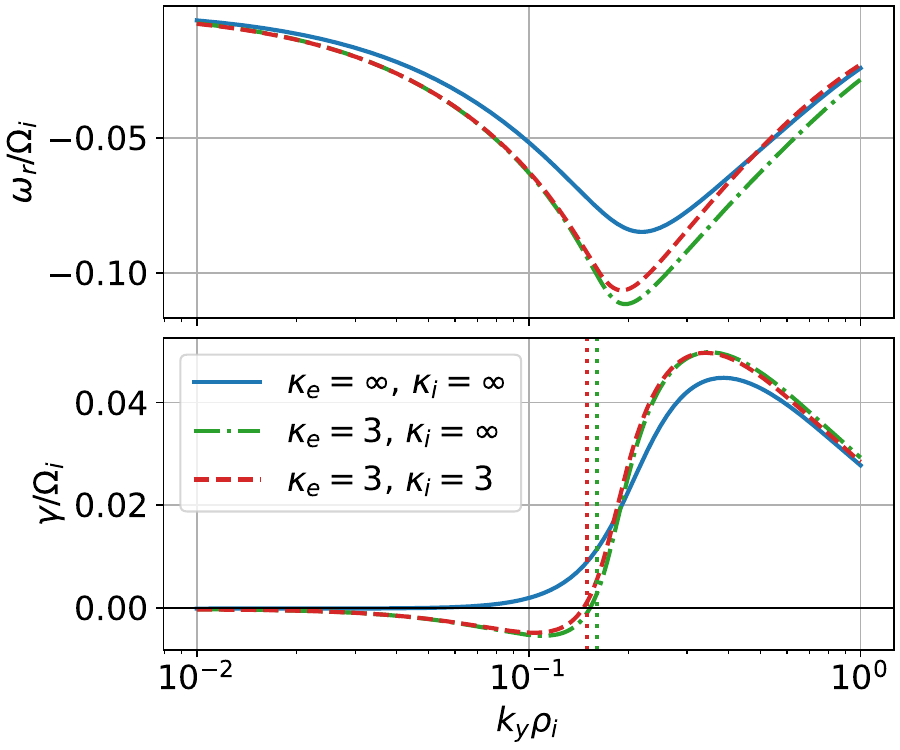}
    \caption{The wave frequency (upper panel) and growth rate (lower panel) of the drift waves in inhomogeneous plasmas.
    The wavenumber in $z$-direction is set as $k_z\rho_i = 10^{-3}$.
    The green dotted line denotes the critical wavenumber $K_y$ from the theoretical prediction \eqref{eq:kyrhoi-cond-2} for $\kappa_e=3$ and $\kappa_i=\infty$,
    and the red dotted line represents that for $\kappa_e=\kappa_i=3$. 
    }
    \label{fig:dr}
\end{figure}

The wave frequency and growth rate of linear drift mode in inhomogeneous Kappa-distributed plasmas are illustrated in Fig. \ref{fig:dr}.
To respectively analyze the suprathermal effects of electrons and ions, we choose three sets of kappa parameters (a) $\kappa_e=\kappa_i=\infty$, (b) $\kappa_e=3$ and $\kappa_i=\infty$, and (c) $\kappa_e=\kappa_i=3$.
Therefore, one can identify the suprathermal effects of electrons by comparing the kappa set (a) with (b), and the suprathermal effects of ions by comparing (b) with (c).
In the upper panel, it shows that the suprathermal electrons increase the absolute value of wave frequency $|\omega_r/\Omega_i|$, while the suprathermal ions take the opposite effect.
In the lower panel, the suprathermal effects of electrons on the growth rate behave distinctly in large and small wavenumbers.
The suprathermal electrons destabilize the wave for roughly $k_y\rho_i > 0.18$ but stabilize
for $k_y \rho_i < 0.18$.
Further, we observe the instability does not occur below some critical wavenumbers when only suprathermalizing the electrons, as we predict in Sec. \ref{sec:appsol}.
The figure also implies that our theoretical threshold wavenumbers \eqref{eq:kyrhoi-cond-2}, plotted by the vertical dotted lines in the lower panel, are close to the exact numerical solutions.
In addition, the suprathermalization of ions plays a less important role than that of electrons; it only affects the growth rate to a small extent.

\begin{figure}
    \centering
    \includegraphics[width=8.5cm]{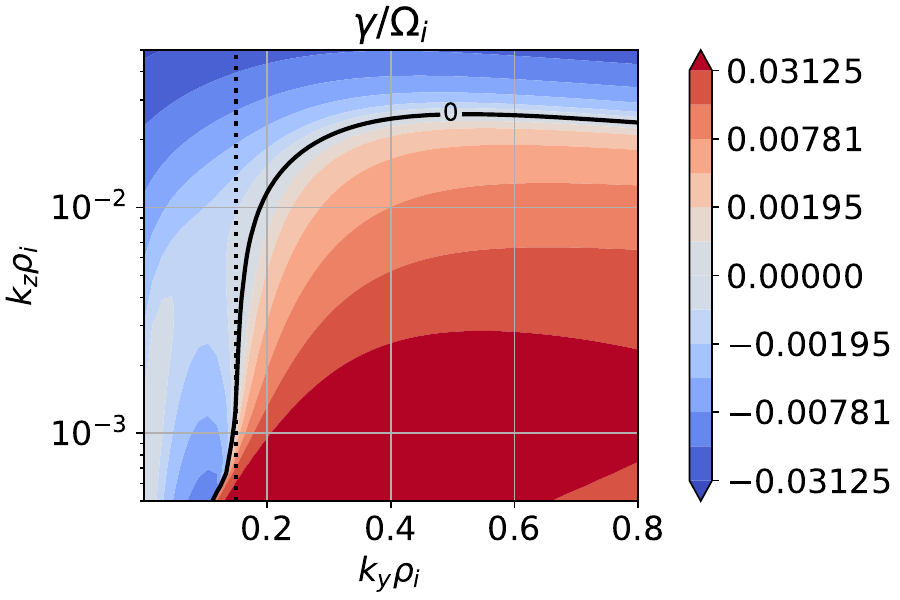}
    \caption{The growth rate $\gamma/\Omega_i$ for varying $k_z\rho_i$ and $k_y\rho_i$ in Kappa-distributed plasmas.
    The kappa parameters are chosen as $\kappa_e=3$ and $\kappa_i=3$.
    The black solid line is the separatrix between stable and unstable drift modes.
    The black dotted line is the theoretical critical wavenumber from Eq. \eqref{eq:kyrhoi-cond-2}.
    }
    \label{fig:gamma_vs_kykz}
\end{figure}
Figure \ref{fig:gamma_vs_kykz} plots the stable and unstable region of $k_y\rho_i$ and $k_z\rho_i$ for drift modes in Kappa-distributed plasmas.
It illustrates that the stable region narrows with a decreased $k_z\rho_i$ but still exists even in very small wavenumber.
The theoretical threshold wavenumber \eqref{eq:kyrhoi-cond-2} is highly accurate around $k_z\rho_i \approx 10^{-3}$ because such wavenumbers coincide with the assumption $\theta_i \ll \omega/k_z \ll \theta_e$ used in the derivations of theoretical $K_y$.

\begin{figure}
    \centering
    \includegraphics[width=8.5cm]{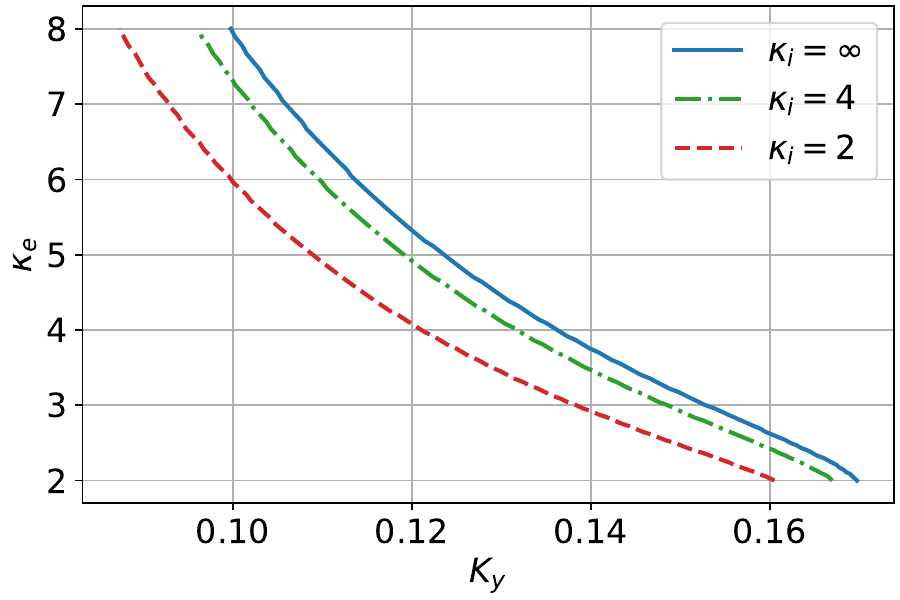}
    \caption{The critical wavenumber $K_y$ numerically solved by the dispersion relation \eqref{eq:dr-kappa-w-lf} for varying $\kappa_e$ and $\kappa_i$.
    The $z$-direction wavenumber $k_z\rho_i=10^{-3}$ is selected.
    }
    \label{fig:kyrhoi_lim}
\end{figure}
Figure \ref{fig:kyrhoi_lim} shows the numerical solutions of the critical wavenumber $K_y$.
It exhibits similar features to the theoretical results.
The critical wavenumbers $K_y$ decreases with an increased $\kappa_e$ or a reduced $\kappa_i$.
For the kappa values from the observations $2<\kappa_{e,i}<8$, the dimensionless wavenumber $K_y$ is a nonzero value and cannot be ignored. 

\subsection{Discussions}
The role of suprathermal electrons in drift instability can be understood as follows.
As we know, the growth rate of drift mode is a combination of the destabilization by the diamagnetic drift of particles and the stabilization by the Landau damping.
As is shown in Sec. \ref{sec:appsol}, the contribution from electrons is much larger than that from ions, so we can only consider the electron contribution.
The strength of electron drift motion is determined by the drift speed $u_e$ or, equivalently, the drift frequency $\omega_{Ne} = k_y u_e$.
With the help of Eq. \eqref{eq:th} and $\rho_\sigma = \theta_\sigma/|\Omega_\sigma|$, one can rewrite the drift frequency of electrons \eqref{eq:drift-freq} in the unit of $\Omega_i$,
\begin{equation}
    \frac{\omega_{Ne}}{\Omega_i} = - k_y \rho_i \frac{\rho_i}{L_N} \frac{m_e}{m_i} \frac{\theta_e^2}{\theta_i^2} \frac{\kappa_e}{\kappa_e-3/2}.
    \label{eq:wne}
\end{equation}
Recalling that $\rho_i/L_N$, $m_e/m_i$, and $\theta_e/\theta_i$ are all constants in numerical analysis, we find that the drift frequency $|\omega_{Ne}/\Omega_i|$ is proportional to $k_y\rho_i \kappa_e/(\kappa_e-3/2)$.
On the one hand, when $k_y \rho_i \ll 1$, the instabilities contributed by the diamagnetic drift are very weak due to the small $|\omega_{Ne}/\Omega_i|$ and can be balanced by the electron Landau damping in Maxwellian plasmas, as shown by the blue solid line in the lower panel of Fig. \ref{fig:dr}.
In Kappa-distributed plasmas, the suprathermal electrons enhance the Landau damping, leading to a more stable drift mode in small wavenumbers.
On the other hand, when $k_y \rho_i$ increases, the diamagnetic drift is strong enough to destabilize the waves.
The suprathermal effects of electrons further enhance the drift motion due to the factor $\kappa_e/(\kappa_e-3/2)$ in $|\omega_{Ne}/\Omega_i|$, resulting in a more unstable drift mode compared with the Maxwellian cases in large wavenumbers, as shown in the lower panel of Fig. \ref{fig:dr}.
Thereby, the combination of the stable mode in small wavenumbers and the unstable mode in large wavenumbers gives rise to the critical wavenumber illustrated in Figs. \ref{fig:dr}-\ref{fig:kyrhoi_lim}.

Furthermore, during the increment of $k_z \rho_i$, the wave propagation deviates from the direction of diamagnetic drift, which weakens the instability.
Therefore, the stable region of drift mode enlarges for large $k_z \rho_i$, as shown in Fig. \ref{fig:gamma_vs_kykz}.

\section{Summary}
\label{sec:sum}
In the present paper, we study the suprathermal effects of electrons and ions on drift instability in inhomogeneous plasmas.
With weak inhomogeneity assumption and local approximation, the linear dispersion relation of drift modes in Kappa-distributed plasmas is derived in a novel integral form \eqref{eq:dr-kappa-w} by utilizing the superstatistics formulas.
This novel approach has the following advantages.
First, the integral representation \eqref{eq:dr-kappa-w} indicates a clear physical meaning that the Kappa dispersion relation can be understood as a weighted superposition of the Maxwellian dispersion relations with fluctuating thermal speeds.
Second, the Kappa dispersion relation \eqref{eq:dr-kappa-w} only includes the standard dispersion function in equilibrium plasmas.
It is needless to define new dispersion functions modified by nonthermal distribution and study their properties. 
Last, this method could be easily extended to other studies on Kappa-distributed plasmas.

Using this new integral representation, we derive the real frequency \eqref{eq:wr} and growth rate \eqref{eq:gamma}-\eqref{eq:gamma_i} in the region of wave speed $\theta_e \ll \omega/k_z \ll \theta_i$.
Unlike the case of Maxwellian plasmas, the unstable drift wave in Kappa-distributed non-uniform plasmas has a lower limit of wavenumber \eqref{eq:kyrhoi-cond-2} in the $y$-direction.
The suprathermalization of electrons is the decisive factor causing such a nonzero critical wavenumber;
the population of suprathermal ions affects the critical wavenumber but does not determine its emergence.
The wave frequency, growth rate, and critical wavenumber are numerically analyzed in Figs. \ref{fig:dr}-\ref{fig:kyrhoi_lim}.
We find that the critical wavenumber decreases with reduced suprathermal electrons or increased suprathermal ions.
Numerical results reveal that such a critical wavenumber cannot be ignored for a typical kappa range $2<\kappa_{e,i}<8$ observed in space plasmas. 
Our study implies that the suprathermal effect of electrons plays a more significant role than that of ions in drift modes.

\begin{acknowledgments}
This work was supported by the National Natural Science Foundation of China (No.12105361) and by the Supporting Fund from Civil Aviation University of China (No.3122022PT18).
\end{acknowledgments}

\section*{Data Availability}
Data sharing is not applicable to this article as no new data were created or analyzed in this study.

\appendix
\section{Properties of weight functions \texorpdfstring{$G_1$}{G1} and \texorpdfstring{$G_2$}{G2}}
\label{sec:g1g2}
\begin{figure}
    \centering
    \includegraphics[width=8.5cm]{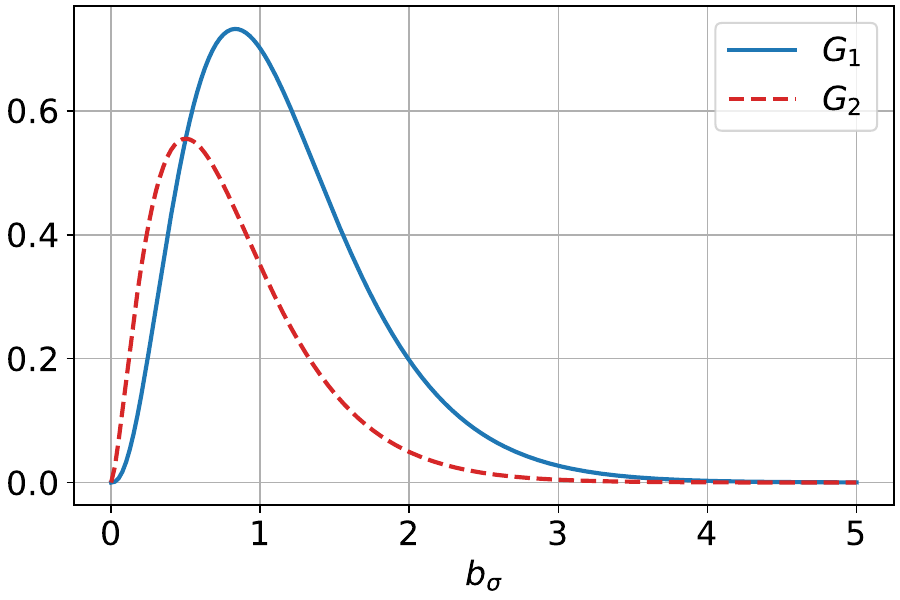}
    \caption{The weight function $G_1(b_\sigma)$ \eqref{eq:G1} and $G_2(b_\sigma)$ \eqref{eq:G2} for $\kappa_\sigma = 3$. }
    \label{fig:g1g2}
\end{figure}

The weight function $G_1(b_\sigma)$ \eqref{eq:G1} is a normalized gamma distribution, while $G_2(b_\sigma)$ \eqref{eq:G2} is an unnormalized gamma distribution.
The plots of these two functions are illustrated in Fig. \ref{fig:g1g2}.
The $n$-order moments of $G_1$ and $G_2$ are, respectively,
\begin{align}
    &\int_0^{+\infty} b_\sigma^n G_1(b_\sigma) \dd{b_\sigma} = \frac{\Gamma(\kappa_\sigma+n+1/2)}{\kappa_\sigma^n \Gamma(\kappa_\sigma+1/2)}, 
    \label{eq:g1-n-moment} \\
    &\int_0^{+\infty} b_\sigma^n G_2(b_\sigma) \dd{b_\sigma} = \frac{(\kappa_\sigma-3/2)\Gamma(\kappa_\sigma+n-1/2)}{\kappa_\sigma^n \Gamma(\kappa_\sigma+1/2)}.
    \label{eq:g2-n-moment}
\end{align}
For $n=0$, Eqs. \eqref{eq:g1-n-moment} and \eqref{eq:g2-n-moment} give the normalizations,
\begin{align}
    &\int_0^{+\infty} G_1(b_\sigma) \dd{b_\sigma} = 1, 
    \label{eq:g1-norm} \\
    &\int_0^{+\infty} G_2(b_\sigma) \dd{b_\sigma} = \frac{\kappa_\sigma-3/2}{\kappa_\sigma-1/2}.
    \label{eq:g2-norm}
\end{align}
In the limit $\kappa_\sigma \rightarrow +\infty$, the $n$-order moments \eqref{eq:g1-n-moment} and \eqref{eq:g2-n-moment} approach,
\begin{align}
    &\lim_{\kappa_\sigma \rightarrow +\infty}\int_0^{+\infty} b_\sigma^n G_1(b_\sigma) \dd{b_\sigma} = 1, 
    \label{eq:g1-n-moment-lim} \\
    &\lim_{\kappa_\sigma \rightarrow +\infty}\int_0^{+\infty} b_\sigma^n G_2(b_\sigma) \dd{b_\sigma} = 1,
    \label{eq:g2-n-moment-lim}
\end{align}
by using the limit of the gamma function, \cite{Olver2010} i.e., $\lim_{z\rightarrow \infty} \Gamma(z+a)/\Gamma(z+b) = z^{a-b}$.
Therefore, if we compute the integral of an arbitrary function $h(b_\sigma)$ weighted by $G_{1}(b_\sigma)$ or $G_{2}(b_\sigma)$, we find,
\begin{align}
    &\lim_{\kappa_\sigma \rightarrow +\infty}\int_0^{+\infty} h(b_\sigma) G_{1,2}(b_\sigma) \dd{b_\sigma} \notag\\
    =& \lim_{\kappa_\sigma \rightarrow +\infty}\int_0^{+\infty} \sum_{n=0}^\infty \frac{h^{(n)}(0)}{n!}b_\sigma^n G_{1,2}(b_\sigma) \dd{b_\sigma} \notag\\
    =&\sum_{n=0}^\infty \frac{h^{(n)}(0)}{n!}\notag\\
    =&h(1).
\end{align}
It implies that both $G_{1}(b_\sigma)$ and $G_{2}(b_\sigma)$ are the delta functions when $\kappa_\sigma \rightarrow +\infty$,
\begin{equation}
    \lim_{\kappa_\sigma \rightarrow +\infty}G_{1,2}(b_\sigma) = \delta(b_\sigma-1).
\end{equation}

\section{Expansion of dispersion relation in the region \texorpdfstring{$\theta_i\ll\omega/k_z\ll\theta_e$}{θi<<ω/kz<<θe}}
\label{sec:expansion}
The dispersion relation expressed in the integral form \eqref{eq:dr-kappa-w-lf} is easy to expand in the wavenumber $\theta_i\ll\omega/k_z\ll\theta_e$ with similar handling of the case in Maxwellian plasmas.
For the electron term, we need to expand the integral,
\begin{equation}
    \int_0^{+\infty} \dd{b_e} 
    \left[G_1(b_e) - G_2(b_e) \frac{\omega_{Ne}}{\omega}\right]
    \left[
    W\left(\xi_e\sqrt{b_e}\right) -1
    \right]
    \Lambda_0 \left(\frac{k_y^2 \rho_e^2}{b_e}\right)
    \label{eq:expansion-e-dev1}
\end{equation}
into the series of $\xi_e = \omega/(k_z\theta_e)$ and then neglect high-order terms $O(\xi_e^2)$.
For this purpose, one expands the dispersion function $W(\xi_e \sqrt{b_e})$ into a series of $\xi_e \sqrt{b_e}$
and then integrates it over $b_e$.
With the well-known expansion, \cite{Ichimaru2004}
\begin{equation}
    W(z) \approx 1 + i \sqrt{\frac{\pi}{2}} z,
\end{equation}
the integral \eqref{eq:expansion-e-dev1} becomes,
\begin{equation}
    i \sqrt{\frac{\pi}{2}} \frac{\omega}{k_z \theta_e}\int_0^{+\infty} \dd{b_e} 
    \left[G_1(b_e) - G_2(b_e) \frac{\omega_{Ne}}{\omega}\right]
    \sqrt{b_e}
    \Lambda_0 \left(\frac{k_y^2 \rho_e^2}{b_e}\right).
    \label{eq:expansion-e-dev2}
\end{equation}
Considering the approximation $k_y \rho_e \ll 1$, one can further expand the above integral \eqref{eq:expansion-e-dev2} into a series of $k_y \rho_e$ and only maintain the zero-order term.
Finally, one finds Eq. \eqref{eq:expansion-e-dev1},
\begin{align}
    & \int_0^{+\infty} \dd{b_e} 
    \left[G_1(b_e) - G_2(b_e) \frac{\omega_{Ne}}{\omega}\right]
    \left[
    W\left(\xi_e\sqrt{b_e}\right) -1
    \right]
    \Lambda_0 \left(\frac{k_y^2 \rho_e^2}{b_e}\right) \notag \\
    \approx & i \sqrt{\frac{\pi}{2}} \frac{\omega}{k_z \theta_e}\int_0^{+\infty} \dd{b_e} 
    \left[G_1(b_e) - G_2(b_e) \frac{\omega_{Ne}}{\omega}\right]
    \sqrt{b_e} \notag \\
    = & i \sqrt{\frac{\pi}{2}} \frac{\omega}{k_z \theta_e}
    \frac{\Gamma(\kappa_e)\sqrt{\kappa_e}}{\Gamma(\kappa_e+1/2)}
    \left(1-\frac{\kappa_e-3/2}{\kappa_e}\frac{\omega_{Ne}}{\omega}\right).
    \label{eq:expansion-e}
\end{align}
For the ion term, with the asymptotic series, \cite{Ichimaru2004}
\begin{equation}
    W(z) \approx -\frac{1}{z^2} + i \sqrt{\frac{\pi}{2}} z e^{-\frac{z^2}{2}},
\end{equation}
the integral can be reduced to,
\begin{align}
    & \int_0^{+\infty} \dd{b_i} 
    \left[G_1(b_i) - G_2(b_i) \frac{\omega_{Ni}}{\omega}\right]
    \left[
    W\left(\xi_i\sqrt{b_i}\right) -1
    \right]
    \Lambda_0 \left(\frac{k_y^2 \rho_i^2}{b_i}\right) \notag \\
    \approx & \int_0^{+\infty} \dd{b_i} 
    \left[G_1(b_i) - G_2(b_i) \frac{\omega_{Ni}}{\omega}\right]
    \left[
    -1+i \sqrt{\frac{\pi}{2}} \xi_i \sqrt{b_i} e^{-\frac{\xi_i^2 b_i}{2}}
    \right]
    \Lambda_0 \left(\frac{k_y^2 \rho_i^2}{b_i}\right), \notag \\
    =& \left(\frac{\omega_{Ni}}{\omega}A_2-A_1\right)
    + i \sqrt{\frac{\pi}{2}} \xi_i \left(B_1-B_2\frac{\omega_{Ni}}{\omega}\right)
    \label{eq:expansion-i}
\end{align}
where $O(\xi_i^{-2})$ is neglected, and the notations $A_1$, $A_2$, $B_1$, and $B_2$ are defined in Eqs. \eqref{eq:A1}-\eqref{eq:B2}.
\bibliography{mylib}
\end{document}